\documentclass[11pt]{article}
\usepackage[utf8]{inputenc}
\usepackage[english]{babel}
\usepackage{overcite,dcolumn,bm,rotating,booktabs}
\usepackage{amsmath,amssymb,latexsym}
\usepackage[normalem]{ulem}
\usepackage[usenames,dvipsnames,svgnames,table]{xcolor}
\usepackage{dcolumn}

\begin{document}

\bibliographystyle{jcc}
\ \\[-7mm]
\begin{center} {\huge \baselineskip=4.0ex 
 Accurate multireference study of Si$_3$ electronic manifold}\\[8mm]  
 
  {\large C. E. M. Gonçalves$^{a}$, B. R. L. Galvão$^{b}$, J. P. Braga$^{a}$ }\\*[6mm] 
 $^{a}${Departamento de Química, Universidade Federal de Minas Gerais, Av. Antônio Carlos 6627, 31270-901, Belo Horizonte, MG, Brazil}\\*[1mm]
  $^{b}${Departamento de Química, Centro Federal de Educação Tecnológica de Minas Gerais, CEFET-MG, Av. Amazonas 5253, 30421-169, Belo Horizonte, Minas Gerais, Brazil}\\*[1mm] 

\vspace{5mm}

(\today)
\end{center}

\baselineskip=5.0ex

\begin{abstract}

Since it has been shown that the silicon trimer has a highly multi-reference character,  accurate multi-reference configuration interaction calculations are performed to elucidate its electronic manifold. 
Emphasis is given to the long range part of the potential, aiming to understand the atom-diatom collisions dynamical aspects, to describe conical intersections and important saddle points along the reactive path. 
Potential energy surface main features analysis are performed for benchmarking, and highly accurate values for structures, vibrational constants and energy gaps are reported, as well as the unpublished spin-orbit coupling magnitude. 
The results predict that inter-system crossings will play an important role in dynamical simulations, specially in triplet state quenching, making the problem of constructing a precise potential energy surface more complicated and multi-layer dependent. 
The ground state is predicted to be the singlet one, but since the singlet-triplet gap is rather small (2.448 kJ/mol) both of them are expected to be populated at room temperature.

\end{abstract}

\section{Introduction}

The interest in understanding properties of small silicon clusters, such as Si$_3$, initially arose in astrophysics researches regarding the carbon-rich stars spectra~\cite{MER26:175,SAN50:262}, and chemical vapor decomposition (CVD) processes~\cite{HO84:51}. Small silicon clusters were also found important as building blocks for semiconductor devices and optoelectronic nanomaterials~\cite{HON93:42,HO98:582}.
Some disilenes synthetic inorganic photochemical reactions also have cyclotrisilanes as starting point\cite{MAS82:1150,MAS83:6524,MUR84:2131}. Therefore, such system is increasing in relevance nowadays, and accurate knowledge of its properties has fundamental importance.

Silicon trimer was first observed in mass spectra\cite{HON54:1610} with small concentrations, and its total atomization energy (TAE) most accurate experimental value is 705$\pm$16 kJ/mol~\cite{SCH95:2574}.
Photoelectron spectrum measured by Kitsopoulos {\em et al.}~\cite{KIT90:6108} showed that its $X$-band resulted from a convolution of features from singlet ($C_{2v}$) and triplet ($D_{3h}$) states. 
These two near degenerates Si$_3$ forms are constantly competing in experiments. For example, Arnold and Neumark~\cite{ARN94:1797} performed threshold photodetachment zero electron kinetic energy spectroscopy on Si$_3^-$, and observed neutral Si$_3$ triplet state vibrational frequencies, but transitions to singlet were not detected. The low-spin form was observed by McCarthy and Thaddeus~\cite{MCC03:213003}, and from recorded rotational spectrum, precise geometrical structure, as an isosceles triangle, were obtained. Further experimental observations~\cite{REI12:194307} recorded the spectrum in 18000-20800 cm$^{-1}$ region,  dominated by triplet-triplet transitions. 
Singlet and triplet states are both important and lie very close together in energy.

The Si$_3$ molecule was also extensively studied by theoretical approaches, such as the Martin and Schaber\cite{MAR79:61} pioneering work. Most {\em ab initio} calculations predict the two lower singlet and triplet states ($^1A_1$ and $^3B_2$) to lie very close in energy, with different methods disagreeing on which state is lower~\cite{GRE85:111,DIE85:29,RAG85:3520,OYE11:094103,TAM15:805}. Vertical and adiabatic electron affinities, as well as vertical ionization energies were first calculated by Niessen and Zakrzewski~\cite{NIE92:1228}, while Kalcher and Sax\cite{KAL98:109,SAX85:469} calculated bending potential energy curves for Si$_3$ under C$_{2v}$ symmetry. 
Further studies were carried out on this system regarding TAE, heat of formation, adiabatic electron affinity, binding and dissociation energy\cite{TAM13:147,AGU14:1397}, total ionization cross-section\cite{NAG13:267}, dynamic hyperpolarizability~\cite{LAN08:118} and photoelectron spectrum \cite{GAR07:074305}.

After recognizing the silicon CVD elementary steps importance, Thompson {\em et al.} carried out trajectories studies to investigate Si$_3$ formation from atom-diatom collisions and silicon dimer formation from three-body thermal recombination, in which cross sections and rate coefficients were obtained. For this purpose, they obtained the first analytical potential energy surface (PES)\cite{MAR86:4426,GAI88:156,MAR90:5311,KAY90:6607} for silicon trimer.

However, the PES for the diatomic case (Si$_2$) is highly complex~\cite{PEY82:111} showing several low lying electronic states (singlet and triplet) with several crossings among them. Therefore, in collisions between Si($^3P$)+Si$_2$, a myriad of states will influence reaction dynamics. This work focus is to describe several Si$_3$ electronic states and their intersections to asses and give insights about dynamical properties of collisions and transitions between electronic states. 
The singlet/triplet states conversion, under ordinary temperatures, is also investigated.
%

\section{Ab initio calculations}

Calculations were carried out at two different theory levels. Geometries were optimized using the quadratic steepest descent method at complete active space self-consistent field (CASSCF) level, with aug-cc-pVQZ (or simply AVQZ) basis set~\cite{DUN89:1007,KEN92:6796}. To improve energy values, each optimized point was recalculated with multi-reference configuration interaction method including Davidson correction MRCI(Q)~\cite{WER88:5803,KNO88:514}, with the same basis. Only singlet and triplet minima had their geometries optimized with MRCI(Q) calculations, for benchmarking purposes. For multiple states calcuations, geometries were optimized only for the lowest state. MOLPRO\cite{molpro} package was used for all calculations.

To improve accuracy, part of the results were extrapolated to complete basis set (CBS) limit employing AV$Q$Z and AV$5$Z basis set, according to USTE method~\cite{VAR07:244105}. 
Extrapolation of raw CASSCF energies was performed with Karton and Martin protocol~\cite{JEN05:267,KAR06:330,VAR07:244105},
\begin{equation}
E_X^{CAS}({\bf R})=E_{\infty}^{CAS}({\bf R})+B/X^{5.34}
\label{cbs1}
\end{equation}
in which subscript $X$ stands for basis sets cardinal number (the CBS limit is represented by $X=\infty$). The variable ${\bf R}$ is a vector representing the molecule's geometry, while $E_{\infty}^{CAS}$ and $B$ are parameters to be determined by fitting the equation to AV$Q$Z and AV$5$Z energies.

The dynamical correlation energy extrapolated to CBS limit ($E_{\infty}^{dc}$) is obtained within USTE method from
\begin{equation}
E_X^{dc}=E_{\infty}^{dc}+\frac{A_3}{(X-3/8)^3}+\frac{A_5^{(0)}+cA_3^n}{(X-3/8)^5}
\label{cbs2}
\end{equation}
in which a fitting process provides $E_{\infty}^{dc}$ and $A_3$. 
Parameters $A_5^{(0)}$, $c$ and $n$ are system-independent and constant for a given post-HF method~\cite{VAR07:244105}. Their values for MRCI method are $A_5^{(0)}$=0.0037685459 $E_h$, $c$=-1.17847713 $E_h^{1-(n)}$ and $n$=1.25~\cite{VAR07:244105,GAL13:4044}.

To assess possible transitions between singlet and triplet states, the magnitude of the spin-orbit coupling (SOC) was also computed. For this purpose, full spin-orbit matrix was calculated using Breit-Pauli operator~\cite{BER00:1823} as implemented in MOLPRO~\cite{molpro} package. In this case, spin-free electronic Hamiltonian eigenstates, labeled $|S\rangle$, $|T,1\rangle$, $|T,0\rangle$ and $|T,-1\rangle$, are used to build the total Hamiltonian matrix representation ($H_{el}+H_{SO}$). The singlet-triplet spin-orbit transition probabilities ($V_{SO}^2$) will depend on

\begin{equation}
V_{SO}^2= \sum_{M_S=-1}^{1} |\langle T,{M_S} | H_{SO} |  S\rangle |^2.
\label{EqVso}
\end{equation}

\section{Results and Discussion}

\subsection{Benchmarks}

Table~\ref{tab:benchmarks} compares results obtained from MRCI(Q)/AVQZ method to experimental data available together with most accurate and recent theoretical results.
The Si$_3$ singlet TAE was measured experimentally by Schmude {\em et al.}~\cite{SCH95:2574} to be 705$\pm$16 kJ/mol.
Oyedepo {\em et al.} obtained a theoretical TAE estimate~\cite{OYE11:094103} employing multireference correlation consistent composite approach~\cite{TEK09:2959} (MR-ccCA), with a result of 710 kJ/mol in excellent agreement with experiments. 
In reference~\citenum{OYE11:094103} it was further noticed that both singlet and triplet forms have strong multirefrence (MR) character. A system with strong MR effects shows a $C_0^2$ (magnitude of the SCF configurations to CASSCF wave functions) value less than 0.90 while $T_1$ and $D_1$ diagnostics~\cite{LEE89:199,LEI00:431,LEE03:362} are larger than the generally accepted cutoff of 0.02 and 0.05. 
The Si$_3$ singlet state, with $C_0^2=0.822$, $T_1=0.032$ and $D_1=0.082$, is therefore not suitably described by a single-reference method by any of those criterion. The triplet state has $C_0^2=0.843$, $T_1=0.031$, $D_1=0.082$, and thus also fails the tests.

\begin{table}
 \begin{center}
 \caption{ {\small Comparison of the present results with experimental and most recent theoretical values for Si$_3$ ground state.$^a$}\medskip}
 \begin{tabular}{cccccc}
\toprule 
			    & Experimental	&Ref.~\citenum{TAM13:147,TAM15:805}	&Ref.~\citenum{OYE11:094103}	& \multicolumn{2}{c}{This work} \\
  \cline{5-6} %
                &				& (SR)$^b$	& (MR)	& AVQZ	& CBS	\\
\hline
\multicolumn{2}{c}{\bf Singlet$^c$}\\
 TAE 			& 705$\pm$16		& 718 		& 710	& 701.48& 719.79	\\
 $R_{SiSi}$ 		& 4.11			& 4.14		& 4.18	& 4.138	& 		\\
 $\theta $ 		& 78.10			& 80.6		& 79.7	& 78.67	& 		\\
 $\omega_{sy}$ 	& 550.6			& 549		&	 	& 548.5	& 		\\
 $\omega_{as}$ 	& 525.1			& 524		& 		& 532.1	& 		\\
 $\omega_b$ 		& 178$\pm$11		& 180		& 		& 186.9	& 		\\*[2mm]

\multicolumn{2}{c}{\bf Triplet$^d$}\\
 S/T Gap			& 4.19			& 0.8		& 0.8	& 2.403	& 2.448	\\

 $R_{SiSi}$		& 				& 4.33		& 4.37	& 4.336	& 		\\
 $\theta $		& 				& 60.0		& 60.0	& 60.00	& 		\\
 $\omega_{sy}$	& 501$\pm$10		& 502		&		& 503.9	& 		\\
 $\omega$		& 337$\pm$10		& 324		&		& 318.6	& 		\\
 $\omega$		& 337$\pm$10		& 325		&		& 318.7	& 		\\*[1mm]

\bottomrule
 \end{tabular} 
 \label{tab:benchmarks} 
 \end{center}
{\small $^a$ Energies are given in kJ/mol, frequencies ($\omega$) in cm$^{-1}$, bond lengths in $a_0$ and bond angles ($\theta$) in degrees}\\
{\small $^b$ Only most accurate coupled cluster result from Tam {\em et al.} for TAE and GAP is given. Geometry was calculated with (U)CCSD(T)/aug-cc-pVQZ and frequencies with CCSD(T)/cc-pV(T+d)Z.}\\
{\small $^c$ Singlet state experimental geometry taken from Ref.~\citenum{MCC03:213003}, while frequencies from references~\citenum{LI95:275,MCC03:213003}}\\
{\small $^d$ Triplet experimental frequencies taken from Ref.~\citenum{REI12:194307}, and S/T gap is an upper bound from NIST Chemistry WebBook~\cite{NIST:WEBBOOK}}
\end{table}

Nevertheless, Tam {\em et al.} reported~\cite{TAM13:147} single reference (SR) coupled cluster calculations employing large aug-cc-pV$X$Z basis set~\cite{DUN89:1007,KEN92:6796} expanded with tight $d$-functions, extrapolated to CBS limit, and including core-valence, scalar relativistic and spin-orbit corrections. 
Their best result~\cite{TAM15:805} of 718 kJ/mol deviates more from experimental result than Oyedepo {\em et al.}~\cite{OYE11:094103}, which may be attributed to this species strong MR character.

In this work, instead of MR composite approach employed by Oyedepo {\em et al.}~\cite{OYE11:094103}, MRCI(Q) calculations were used, extrapolated to CBS limit. Specifically for obtaining a TAE accurate value, MRCI/AVQZ geometry optimization was performed, followed by energy refinement by extrapolating it with Q,5 pair. Frequencies and zero point energies (ZPE) were evaluated at CASSCF/AVQZ level. Results are given in Table~\ref{tab:benchmarks}, compared with those of Oyedepo {\em et al.}~\cite{OYE11:094103} and Tam {\em et al.}~\cite{TAM13:147,TAM15:805}. The present calculation indicates a TAE value of 719.79 kJ/mol (including ZPE). This value is similar to Tam {\em et al.}~\cite{TAM13:147,TAM15:805}, which takes into account smaller effects, such as core-valence and scalar relativistic corrections. In fact, reference~\citenum{TAM15:805} shows that these smaller corrections add up only to 0.215 kJ/mol, and thus are not the most important factor.

Both singlet and triplet species lie very close in energy, and their gap is still a controversial subject. NIST Chemistry WebBook~\cite{NIST:WEBBOOK} provides an upper bound of 4.19 kJ/mol for this gap. Employing dozens of different calculation methods, Tam {\em et al.}~\cite{TAM15:805} showed that most DFT functionals predicts triplet as the ground state, while most wave function based methods give the singlet state, although exceptions exist. Their best estimate indicates a gap of only 0.8 kJ/mol in singlet favor. The only MR calculation performed in their work was CASSCF and CASPT2, both showing poor results. 
However, the composite scheme of Oyedepo {\em et al.}~\cite{OYE11:094103} agrees with their best prediction as shown in Table~\ref{tab:benchmarks}.
The MRCI results for the gap led 2.40 kJ/mol in AVQZ basis and 2.45 at CBS limit, which differs considerably from results of references~\citenum{OYE11:094103,TAM15:805}, even though somewhat closer to NIST's reported value.

\subsection{Atom-diatom interactions}

\begin{figure}[h]
 \centering \includegraphics[width=9cm,angle=-90]{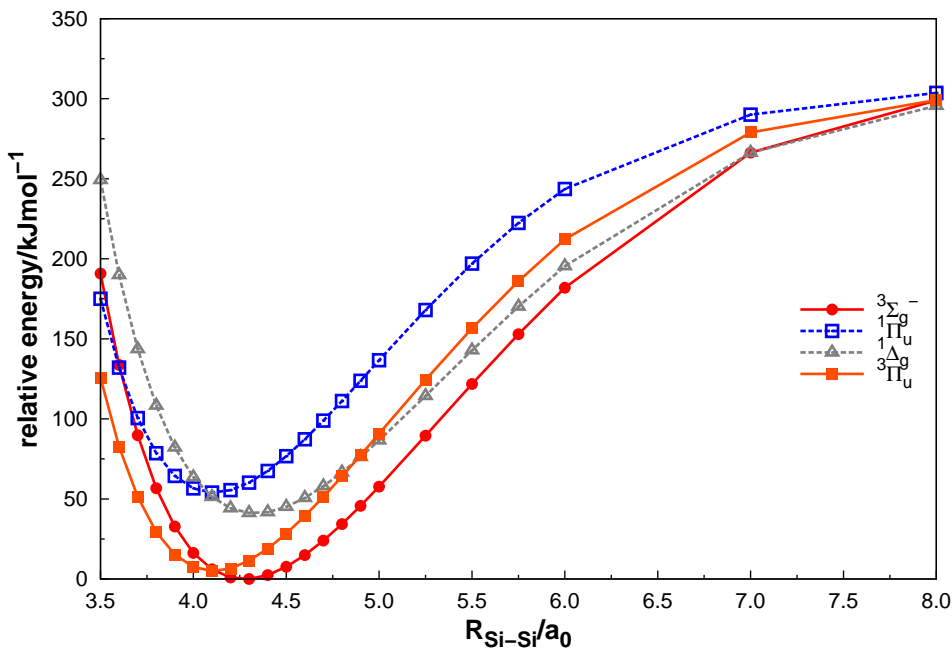}
\caption{{\small Low lying electronic states of Si$_2$. Triplet states are shown with solid lines, while {\bf singlet states} in dashed lines. }}
  \label{fig:si2}
\end{figure}

If the interest is reactive collisions that lead to Si$_3$, such as atom-diatom channel or the termolecular path, there will be a myriad of electronic states involved. For example, focusing only in the Si$_2$ diatom, there are two nearly degenerate states ($^3\Pi_u$ and $^3\Sigma_g^-$), as shown in figure~\ref{fig:si2}. They alternate as ground state for different bond lengths, and it can be expected that both levels will be populated at normal temperatures. Therefore, a collision with a Si($^3$P) atom may happen in either of them, and will unfold an even larger number of electronic states on Si$_3$ molecule. Figure~\ref{fig:si2} also shows other low singlet and triplet states, which may not be accessible under normal experimental circumstances, but are still below dissociation limit to two separate Si($^3$P) atoms.

\begin{figure}[h]
 \centering \includegraphics[width=9cm,angle=-90]{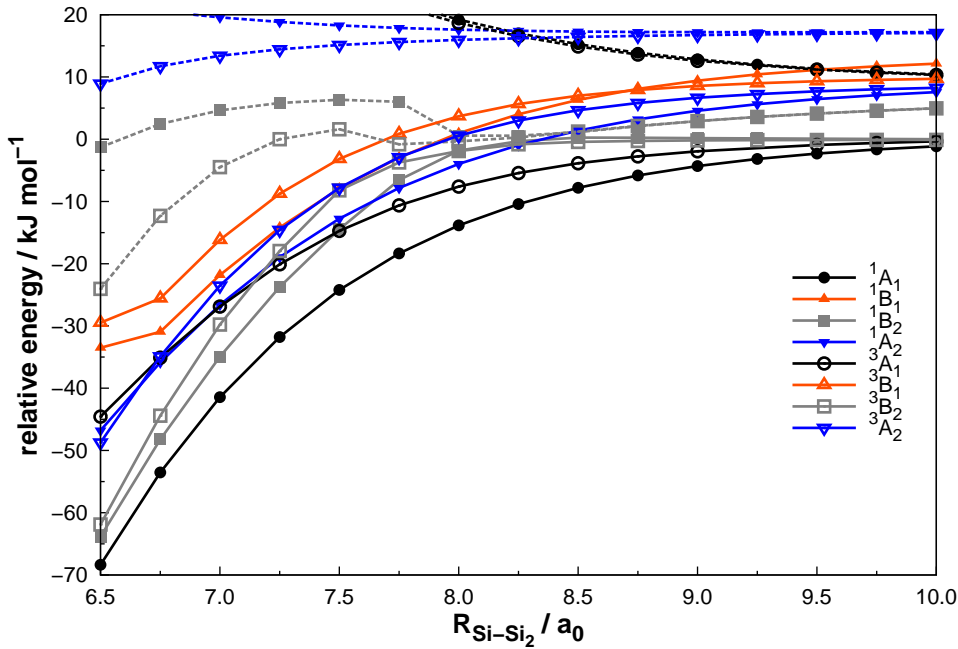}
\caption{{\small T-shaped atom-diatom interaction profile with Si$_2$ separation fixed at 4.5 a$_0$. Two-state calculations were performed under $A_1$, $B_1$, $A_2$, $B_2$ symmetries both for singlet and triplet multiplicities at MRCI(Q)/AVQZ level. Solid lines represents ground states of each symmetry, while dashed lines represent excited ones.}}
  \label{fig:longprofile}
\end{figure}

The analysis of SI$_3$ starts with the construction of potential energy curves describing a T-shaped attack, in which a silicon atom approaches a Si$_2$ molecule (with bond length fixed at 4.5 $a_0$). Figure~\ref{fig:longprofile} points were obtained from two-states calculations under $C_{2v}$ symmetry on each symmetry ($A_1$, $B_1$, $A_2$, $B_2$), both for singlet and triplets states, yielding 16 electronic potential curves. There are several attractive states that lead to a barrierless collision, and thus many of them may contribute to cross sections and rate constants of reactive events. This large amount of states makes the construction of global potential energy surfaces for dynamics studies a very complicated problem. Crossings between singlet and triplet states are also observed, and their coupling are described in the next section. 
Under three-state calculations, even more low lying states may appear, but not as attractive. As can be guessed from figure~\ref{fig:si2}, if such energy profile is repeated for different values of Si$_2$ bond length, a different order may be obtained at the asymptotic limit.

\begin{figure}[h]
 \centering \includegraphics[width=9cm,angle=-90,bb=63 50 541 742 ]{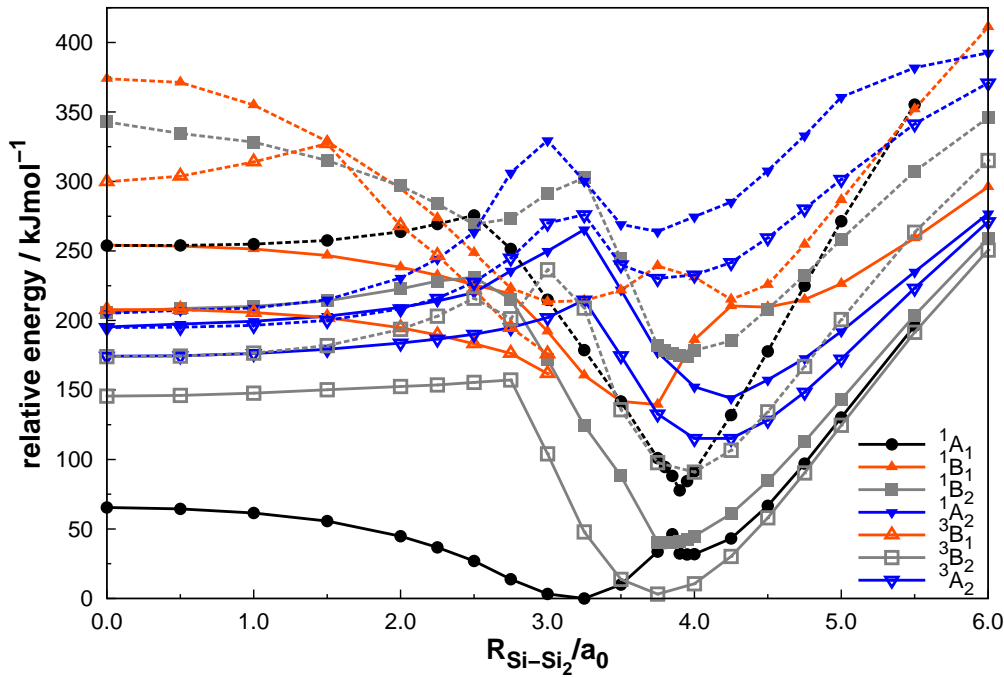}
\caption{{\small Potential curves for T-shaped MEP at MRCI(Q)/AVQZ two-state level. Excited states of a given symmetry are shown with dashed lines. Due to convergence problems, only the triplet $B_2$ and $A_2$ states are shown.}}
  \label{fig:tshaped}
\end{figure}

To explore also the PES covalent region and to have a clearer idea about how far in energy are the excited states situated, two-states calculations were performed under $C_{2v}$ point group for $A_1$, $B_1$, $A_2$, $B_2$ symmetries to describe a T-shaped attack minimum energy path (MEP). For each value of the separation between a Si atom and the Si$_2$ center of mass, the diatom bond length was allowed to relax in CASSCF/AVQZ level, with lower state as reference for optimizations. Each point energy was refined with MRCI(Q)/AVQZ calculation. Therefore, at the right hand side of the graph dissociation limit is approached, and at the left-hand side, with a separation equal to zero, lies linear structure of Si$_3$. Results are shown in figure~\ref{fig:tshaped}.
Singlet ($C_{2v}$) and triplet ($D_{3h}$) minima was observed together with their crossing. Although these two low-lying states show high vertical excitation energies, all excited states are close in energy, indicating several crossings between them. As Si-Si$_2$ limit is approached, this mixing of states is specially pronounced.
 
\begin{figure}[h]
 \centering \includegraphics[width=9cm,angle=-90]{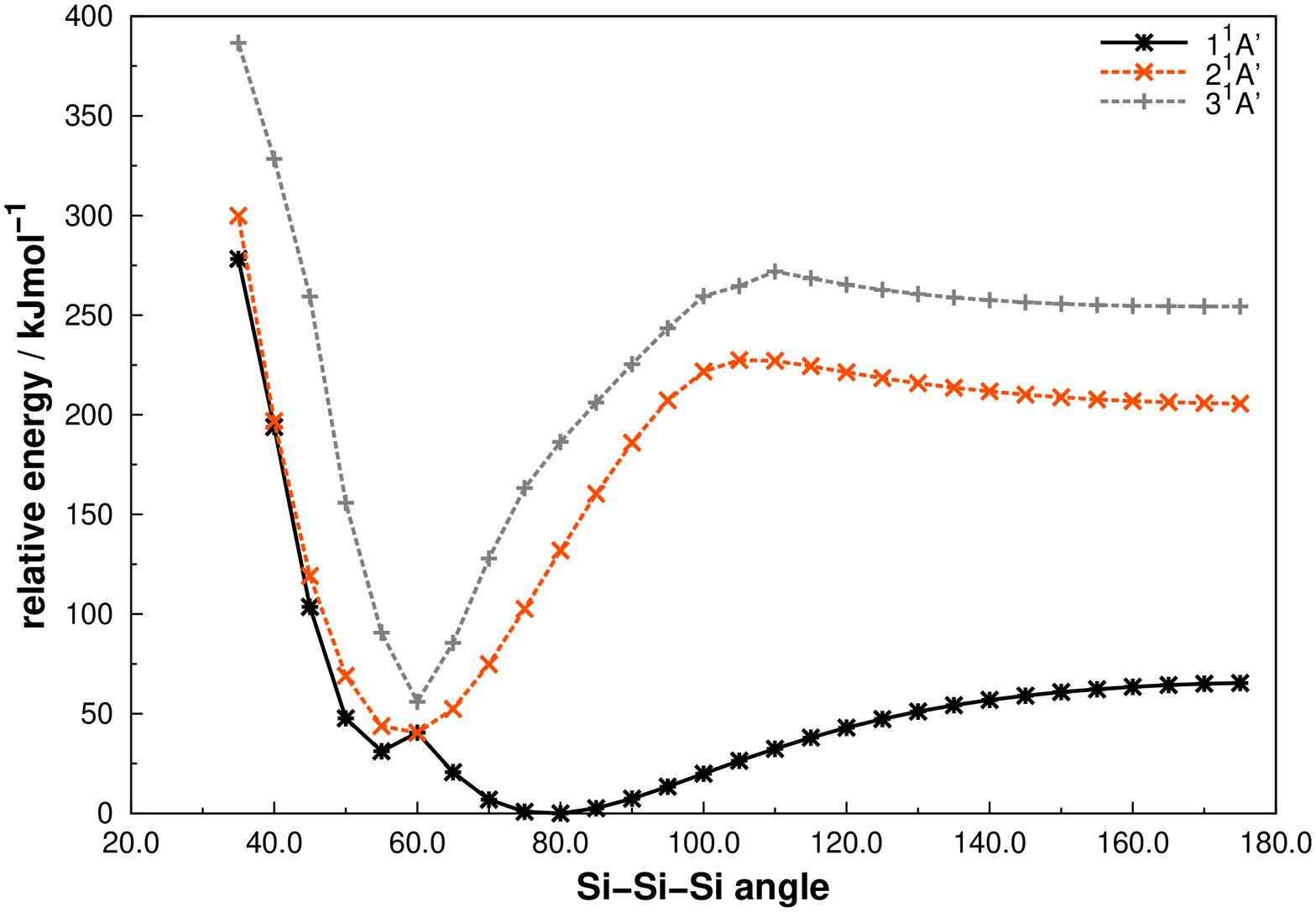}
\caption{{\small Potential curve under $C_s$ symmetry for the three lowest singlet states. }}
  \label{fig:cs-al-3state}
\end{figure}

At $D_{3h}$ configuration, in which triplet state shows a minimum, three singlet states become nearly degenerate \cite{GAR07:074305}. 
Figure~\ref{fig:cs-al-3state} gives a better perspective of this phenomena, in which three-state MRCI(Q) calculations under $C_s$ symmetry were made, but this time varying the bend angle and letting the Si bond distances free to optimize. The two lowest states correspond to the $^1A_1$ and $^1B_2$ first states, and the third is $^1A_1$ second state. The intersection geometry is equilateral, with a bond length of 4.437 a$_0$ and 52.12 kJ/mol above ground state, as shown in figure~\ref{fig:cs-al-3state}.

\begin{figure}[h]
 \centering \includegraphics[width=9cm,angle=-90]{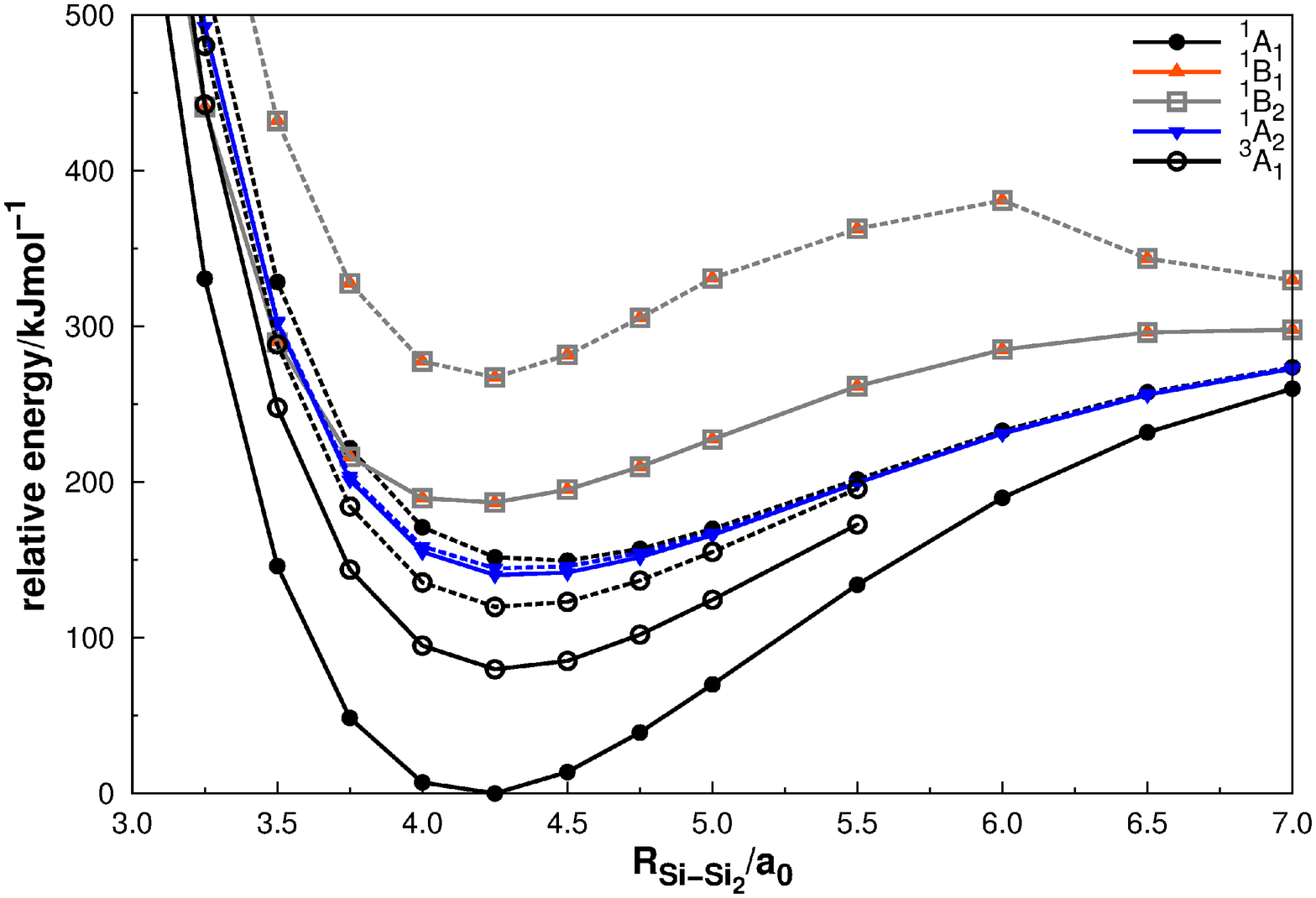}
\caption{{\small Linear profile of Si$_3$ system in CASSCF/AVQZ level. Dashed lines stands for respectives first excited states.}}
  \label{fig:linear}
\end{figure}

The Si-Si$_2$ approach via linear configurations were also investigated, in which a $C_{\infty v}$ minimum energy path is obtained by optimizing Si$_2$ bond length for several Si-Si$_2$ distances, at CASSCF/AVQZ level.
Due to convergence problems, some high lying triplet states could not be presented and only $^3A_1$ converged few points. Results are seen in figure~\ref{fig:linear}. Note that calculations were performed under the $C_{2v}$ symmetry. 
Therefore, the labels $A_1$, $B_1$, $A_2$, $B_2$ were used, but they may not correspond to those in figure~\ref{fig:tshaped}, since the symmetry axis for linear arrangements is different. The minima in these curves may have an imaginary frequency in bending mode, which may be distinguished by analyzing figure~\ref{fig:tshaped}. At the two lowest minima the molecule is symmetric, having equal bond lengths: 4.193 a$_0$ for $^1A_1$ state and 4.290 a$_0$ for $^3A_1$.
The linear dissociation path does not show as many conical intersections as in T-shaped case.

\subsection{Intersystem Crossing and Spin-Orbit Coupling}

Recently, Tam \textit{et al.}~\cite{TAM15:805} explored singlet-triplet gap with several different methods and basis sets, obtaining very different results depending on the calculations level of accuracy (varying from -17.5 to 19.2 kJ/mol). They concluded that the two states can be considered degenerated, with their most accurate calculations yielding a gap of 0.8 kJ/mol in singlet favor. For singlet-triplet crossing point (STCP), they have optimized independently each state for different bond angles, and found that the two states attain similar energy at a bending angle of 68$\pm 2\,^{\circ}$, and this point is 16$\pm$4 kJ/mol above the ground state. In principle, however, the minimum on seam of crossing could be located in a distorted $C_s$ geometry, and thus the location of this STCP was fully optimized to verify this possibility. This was performed at CASSCF level, but employing TZVPP basis set~\cite{WEI98:143}.
It is confirmed that the minimum on the crossing seam lies indeed in a $C_{2v}$ geometry, with a bond length of 4.282 bohr and an angle of 65.7$\,^{\circ}$. To refine CASSCF values, a fine grid of MRCI(Q)/TZVPP energies (now considering only $C_{2v}$ arrangements) was calculated and a bond length of 4.230 bohr and an angle of 67.0$\,^{\circ}$ was obtained. 

Note that the triplet minimum geometry lies only 7.0$\,^{\circ}$ apart from STCP, and the energy difference between them is only 5.75 kJ/mol at MRCI(Q)/TZVPP level (considering ZPE). This energy can be gained by triplet state upon excitation of a single quantum in the bending mode, which implies that intersystem crossings may play a very important role.  From singlet species, the crossing lies 7.58 kJ/mol. Although this is also a small difference, and intersystem crossings may also happen, more bending mode excitation is necessary, and the geometry is also further apart.

\begin{figure}[h]
 \centering \includegraphics[width=9cm,angle=-90]{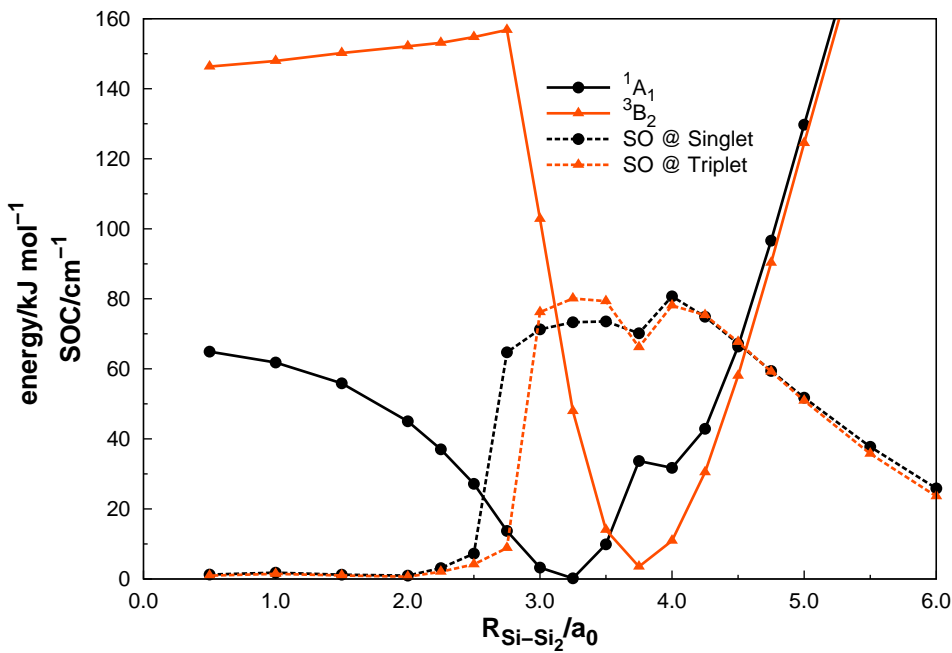}
\caption{{\small Spin-Orbit coupling between $^1A_1$ and $^3B_2$ states. Black and orange dashed lines are for coupling between the two states using singlet and triplet MEP geometries, respectively.}}
  \label{fig:so}
\end{figure}

Given that quenching of triplet state to singlet could imply in significant consequences for Si$_3$ spectroscopic observation, and that there is no reported information regarding the two states coupling magnitude in the literature, MRCI(Q)/AVQZ spin-orbit coupling calculations were performed, and results shown in figure~\ref{fig:so}. SOC has been calculated over the singlet and triplet MEP, with similar results. As seen in this figure, at STCP region the SOC value is about 75 cm$^{-1}$.

By a close observation of figure~\ref{fig:so}, MRCI(Q)/AVQZ prediction of STCP can be estimated and compared to those reported above, with TZVPP basis set (which employed more rigorous search methods). Relatively to singlet state, the crossing lies 6.12 kJ/mol above it (considering ZPE), while for triplet state the crossing point is only 3.71 kJ/mol above the ZPE, which is even lower than what was predicted by TZVPP calculation (with a more rigorous searching method using AVQZ basis set these values would decrease). This indicates that AVQZ basis set predicts even more mixing.
Therefore, all this work's accurate multi-reference calculations predict that intersystem crossings will be very important to this system.

\newpage

\section{Conclusions}

The Si$_3$ molecule was revisited with the powerful MRCI(Q) method, since it had been shown that this system has a strong MR character~\cite{OYE11:094103}. As reported before, singlet and triplet states possess very similar energies, and our calculations showed a singlet-triplet gap of 2.448 kJ/mol (CBS value) favouring the low-spin state. Values of TAE, singlet and triplet frequencies and geometries were computed for benchmarking purposes. Analysing Si$_3$ potential energy surfaces up to the dissociation limits, it becomes clear that collisions in this system will be highly complex, leading to many electronic states with several crossings between them. 
The first Si$_3$ linear study is reported here, which has fewer crossings if compared to C$_{2v}$ geometries, lying more than 50kJ/mol above Si$_3$ minimum. Also unpublished so far, the spin-orbit coupling magnitude between singlet and triplet states are reported, using rigorous MRCI(Q) method which provided a MR prediction of the crossing energy. And above all, results indicates that intersystem crossings will be very important to this system, specially to triplet-singlet quenching.

{\allowdisplaybreaks
\section*{Acknowledgments} 
The authors would like to thank CAPES, FAPEMIG and CNPq for the financial support.
}
\bibliography{bibliography}

\end{document}